%% file: LBK-Mauscript.tex
\documentclass[manuscript]{acmart}
\usepackage{tabularx} 
\usepackage{hyperref}
\usepackage{balance}
\usepackage{enumitem} 
\usepackage{makecell}
\usepackage{graphicx}
\usepackage{subcaption}

\AtBeginDocument{%
  }

\setcopyright{acmlicensed}
\copyrightyear{2025}
\acmYear{2025}
\setcopyright{rightsretained}
\acmConference[CHI EA '25]{Extended Abstracts of the CHI Conference on Human Factors in Computing Systems}{April 26-May 1, 2025}{Yokohama, Japan}
\acmBooktitle{Extended Abstracts of the CHI Conference on Human Factors in Computing Systems (CHI EA '25), April 26-May 1, 2025, Yokohama, Japan}\acmDOI{10.1145/3706599.3719734}
\acmISBN{979-8-4007-1395-8/2025/04}

\begin{document}

\title{Measuring User Experience Through Speech Analysis: Insights from HCI Interviews}


\author{Yong Ma}
\affiliation{%
  \institution{University of Bergen}
  \city{Bergen}
  \country{Norway}}
  \email{yong.ma@uib.no}

\author{Xuedong Zhang}
\orcid{0009-0002-6361-1717}
\affiliation{%
\institution{LMU Munich}
  \city{Munich}
  \country{Germany}}
\email{xuedong.zhang@ifi.lmu.de}

\author{Yuchong Zhang}
\email{yuchongz@kth.se}
\affiliation{%
  \institution{KTH Royal Institute of Technology}
  \city{Stockholm}
  \country{Sweden}
}

\author{Morten Fjeld}
\affiliation{%
\institution{University of Bergen}
  \city{Bergen}
  \country{Norway}}
\email{morten.fjeld@uib.no}
\additionalaffiliation{%
\institution{Chalmers University of Technology}
  \city{Gothenburg}
  \country{Sweden}}
\email{fjeld@chalmers.se}

\renewcommand{\shortauthors}{Ma and Zhang et al.}

\begin{abstract}
User satisfaction plays a crucial role in user experience (UX) evaluation. Traditionally, UX measurements are based on subjective scales, such as questionnaires. However, these evaluations may suffer from subjective bias. In this paper, we explore the acoustic and prosodic features of speech to differentiate between positive and neutral UX during interactive sessions. By analyzing speech features such as root-mean-square (RMS), zero-crossing rate(ZCR), jitter, and shimmer, we identified significant differences between the positive and neutral user groups. In addition, social speech features such as activity and engagement also show notable variations between these groups. Our findings underscore the potential of speech analysis as an objective and reliable tool for UX measurement, contributing to more robust and bias-resistant evaluation methodologies. This work offers a novel approach to integrating speech features into UX evaluation and opens avenues for further research in HCI. 
\end{abstract}




\maketitle

\input{Sections/Introduction}
\input{Sections/Related_Work}
\input{Sections/Experiment_Setup}
\input{Sections/Results}

\input{Sections/Discussion}
\input{Sections/Conclusion}

\begin{acks}
 This research was partially supported by the Research Council of Norway under project number 326907. It was also funded by the Deutsche Forschungsgemeinschaft (DFG, German Research Foundation) -- SPP 2199 -- in the Project TRANSFORM.
\end{acks}

\bibliographystyle{ACM-Reference-Format}
\bibliography{LBK-Mauscript}

\appendix
\renewcommand{\thetable}{A\arabic{table}} 
\clearpage
\onecolumn 

\section{The Result of User Satisfaction Classification}

\begin{table*}[ht]
\centering
\small 
\renewcommand{\arraystretch}{1.3} 
\setlength{\tabcolsep}{6pt} 
\caption{User Satisfaction Classification Using LLM-Based Sentiment Analysis and NLP}
\label{tab:analysis}
\begin{tabular}{@{} c c c c c c @{}}
\toprule
\textbf{Participant} & 
\textbf{SART Accuracy} & 
\textbf{SART Reaction Time} & 
\textbf{AFI Score} & 
\textbf{Feedback Summary} & 
\textbf{Satisfaction} \\
 & (Before → After) & (Before → After) & Improvement &  &  \\ 
\midrule
P1  & 80\% → 80\%  & 0.43 → 0.41  & 72 → 108 (+36)  & Positive, despite safety concerns. Liked VR.  & Positive  \\
P2  & 90\% → 90\%  & 0.50 → 0.53  & 86 → 113 (+27)  & Liked relaxation but music was repetitive.  & Positive  \\
P3  & 100\% → 80\%  & 0.48 → 0.50  & 65 → 89 (+24)  & Concept was good but wanted more engagement.  & Neutral  \\
P4  & 70\% → 90\%  & 0.45 → 0.44  & 36 → 107 (+71)  & Understood concept, relaxed; suggested more studies.  & Positive  \\
P5  & 70\% → 60\%  & 0.42 → 0.37  & 25 → 93 (+68)  & Relaxing but safety concerns while driving.  & Neutral  \\
P6  & 100\% → 50\%  & 0.48 → 0.46  & 51 → 88 (+37)  & Liked the idea but lacked immersion.  & Neutral  \\
P7  & 50\% → 40\%  & 0.47 → 0.46  & 31 → 106 (+75)  & Wanted more customization options.  & Neutral  \\
P8  & 60\% → 90\%  & 0.41 → 0.42  & 59 → 97 (+38)  & Positive experience but requested more music.  & Positive  \\
P9  & 10\% → 60\%  & 0.41 → 0.46  & 68 → 92 (+24)  & Useful but headset felt uncomfortable.  & Neutral  \\
P10 & 80\% → 80\%  & 0.45 → 0.42  & 56 → 110 (+54)  & Liked the system but device was heavy.  & Positive  \\
P11 & 70\% → 60\%  & 0.42 → 0.39  & 43 → 89 (+46)  & Wanted seamless music playback.  & Neutral  \\
P12 & 40\% → 70\%  & 0.41 → 0.46  & 47 → 91 (+44)  & Liked concept but suggested better hardware.  & Neutral  \\
\bottomrule
\end{tabular}
\end{table*}

\twocolumn 

\end{document}

%% file: Sections/Introduction.tex
\section{Introduction}
User satisfaction is a fundamental aspect of evaluating user experience (UX), directly influencing the success of interactive systems and applications~\cite{law2009understanding}. Traditionally, UX evaluations rely on subjective tools, such as questionnaires~\cite{hinderks2019developing,zhang2023industrial} and self-reported scales~\cite{graser2023quantifying}. While these methods provide valuable insights, they are prone to subjective bias, which compromises the consistency and objectivity of the results. These limitations hinder the effective evaluation of UX, particularly user satisfaction, in complex and dynamic real-world scenarios. Such biases can also impede the iterative design process, slowing progress toward enhancing user-centered design and overall system usability.

To address the limitations of questionnaires, user interviews have emerged as an alternative approach for UX evaluation. Compared to questionnaires, interviews allow for a more detailed exploration of user behaviors, emotions, and motivations, providing rich qualitative data~\cite{portigal2023interviewing}. They enable researchers to observe how users interact with systems and to evaluate UX in real-life settings. When coupled with sentiment analysis~\cite{rahman2020sentiment,zhai2022study} and natural language processing (NLP)~\cite{jang2022satisfied}, interview scripts can provide qualitative insights into user satisfaction, usability, and other UX metrics. However, qualitative and textual methods alone may still be influenced by subjective bias, leaving room for more objective evaluation techniques.

Speech analysis offers a promising solution to this challenge. Speech contains rich acoustic and prosodic information, including emotional cues~\cite{george2024review} and social speech features~\cite{ma2009secret,soman2010social}. These characteristics make it an ideal medium for objectively assessing user satisfaction. By evaluating metrics such as short-time energy, zero-crossing rate~\cite{bachu2008separation,ma2021you}, jitter, and shimmer~\cite{li2007stress}, along with social features like engagement and activity~\cite{soman2010social,pentland2007social}, it is possible to distinguish between positive and neutral UX experiences more reliably. This approach can complement traditional methods by addressing the biases inherent in subjective tools.

Speech-based UX evaluation has emerged as a promising objective alternative to traditional self-reported measures. Prior research~\cite{fan2021older} has demonstrated the effectiveness of analyzing speech patterns to infer user engagement and emotional responses in older adults through think-aloud verbalizations. Their study highlighted how acoustic and prosodic features could uncover usability issues and emotional states, providing a strong foundation for the use of speech analysis in UX evaluation. Building on this foundation, our study explores the potential of acoustic and prosodic speech features as reliable indicators for UX evaluation in a broader context. Specifically, we examine short-time energy, zero-crossing rate, jitter, shimmer, engagement, and activity to differentiate between positive and neutral user experiences. Our findings highlight the capability of speech analysis to detect significant differences between user groups, offering a robust and bias-resistant approach to UX assessment. This work introduces a novel methodology for integrating speech-based analysis into UX evaluation, contributing to advancements in HCI research and paving the way for future innovations in adaptive, user-centered system design.

%% file: Sections/Related_Work.tex
\section{RELATED WORK}

\subsection{UX Measurement}
UX evaluation has traditionally relied on subjective tools such as UX questionnaires~\cite{melin2020questionnaire} and self-reported scales~\cite{graser2023quantifying} to assess satisfaction, usability, and engagement. Popular frameworks like the System Usability Scale (SUS) and AttrakDiff provide structured approaches to capturing user perceptions~\cite{hinderks2019developing}. While these tools offer valuable insights, they are often limited by subjectivity bias, making it challenging to derive consistent and objective results~\cite{law2014attitudes}.
This limitation has driven researchers to explore alternative approaches to achieve a deeper understanding of user experiences while minimizing subjectivity bias. One such alternative is behavioral measurement, which focuses on analyzing user behavior by collecting data on actions and responses. This method employs objective metrics such as task completion time, success rate, error rate, and eye movement tracking~\cite{mortazavi2024exploring}. By emphasizing what users actually do, rather than relying solely on self-reported experiences, behavioral measurement offers a more objective perspective on UX evaluation. Another promising alternative is physiological measurement, which examines the relationship between UX and physiological signals such as electroencephalograms (EEGs), heart rate, and pupil dilation~\cite{zhang2024mind, zaki2021neurological, mortazavi2024exploring}. For example, Holman and Adebesin~\cite{10.1145/3351108.3351139} reduced subjectivity in UX measurement by utilizing EEGs to identify brain activity associated with key UX metrics, such as emotions, engagement, and relaxation.

Recently, the impressive capabilities of Large Language Models (LLMs) in natural language understanding, knowledge inference, and contextual semantic processing have attracted significant attention in UX research. Graser et al.\cite{graser2024using} highlighted the inconsistencies in naming and measuring UX factors in traditional questionnaires. To address this, they used ChatGPT-4 to identify semantically similar items and cluster topics from 40 established UX questionnaires, aiming to standardize the description of UX dimensions. Zheng et al.\cite{zheng2024evalignux} introduced EvAlignUX, an LLM-powered system designed to assist HCI researchers in developing UX measurement plans and mapping the relationships between UX metrics and research outcomes. Hsueh et al.~\cite{hsueh2024applying} developed an automated UX testing tool leveraging LLMs to tackle challenges such as subjectivity and lack of standardization in traditional UX measurement methods.
These advancements underscore the potential of alternative methodologies and AI-powered tools in mitigating biases and improving the reliability of UX evaluation.

\subsection{Speech analysis in UX Measurement}
The human voice conveys a wealth of information beyond basic demographic details such as gender and age, encoding valuable cues about emotions~\cite{gangamohan2016}, intentions~\cite{dighe2023audio}, and overall cognitive state~\cite{elbanna2022hybrid, yamada2023smartwatch}. Speech analysis, defined as the process of extracting meaningful information from speech signals~\cite{Toledano2009}, can be categorized into three primary domains. Acoustic analysis focuses on features such as pitch, intensity, and spectral patterns, which are used to identify emotions~\cite{yildirim2004acoustic}, speaker identity~\cite{de2007sound,sun2017unsupervised}, and health-related conditions~\cite{wang2024sound}. Paralinguistic analysis emphasizes non-verbal components, including tone, prosody, and speech rate, providing insights into emotional states~\cite{phukan2024paralingual,ma2023emotion}, intentions~\cite{9383505}, and levels of engagement~\cite{10.1145/2818346.2820775, haider2015high}. Finally, linguistic analysis examines the meaning and structure of speech to enable applications such as speech-to-text conversion~\cite{sethiya2024end}, speech synthesis~\cite{ning2019review}, and dialogue systems~\cite{10.1145/3166054.3166058}. These three domains collectively drive advancements in areas such as emotion recognition, human-computer interaction, and health diagnostics.

Additionally, social speech features, such as speech activity, engagement, emphasis, and mirroring, etc.~\cite{curhan2007thin,pentland2007social,pentland2010honest}, provide valuable insights into conversational dynamics and interpersonal communication, making them relevant for UX evaluation. Speech activity measures the amount of time a speaker is talking, reflecting their participation and involvement~\cite{curhan2007thin,soman2010social}. Engagement evaluates how a speaker's contributions influence the flow of conversation, often linked to turn-taking patterns and dynamic interactions, with higher engagement levels indicating smoother and more interactive exchanges~\cite{pentland2007social,soman2010social}. Emphasis, which refers to variations in vocal stress or loudness, conveys emotional intensity and helps highlight key points in a conversation, maintaining attention~\cite{pentland2010honest,soman2010social}. Mirroring, the synchronization of speech patterns such as pitch and tempo between speakers, is associated with rapport and mutual understanding~\cite{curhan2007thin,soman2010social}. These social speech features can be extensively studied in HCI research for their ability to capture subtle social behaviors that are related to user satisfaction. 

%% file: Sections/Experiment_Setup.tex
\section{Experiment Setup} 
We conducted a user study on in-car virtual natural environments (VRE), developed as part of a separate project~\cite{li2021journey,ma2021you}, to explore the potential of speech analysis for UX measurement. By analyzing post-VRE user interviews, we evaluated the restorative effects and demonstrated the feasibility of using speech analysis for UX evaluation.

\subsection{Apparatus, Participants and Experimental Procedure}
\subsubsection{Apparatus and Participants}
The prototype used in this study was developed using Unity 3D and deployed on a standard PC connected to an Oculus Rift headset, two tracking sensors, and a touch controller. Audio output was delivered through Bluetooth noise-canceling headphones. A tablet was employed for administering questionnaires, and a digital recorder was used to capture participants' voices during post-experience interviews. The recorder was configured to operate at a sampling rate of 48 kHz and 16-bit resolution, with the microphone positioned approximately 15 cm from the participant. For speech analysis, we utilized ELAN\footnote{\href{https://archive.mpi.nl/tla/elan}{https://archive.mpi.nl/tla/elan}} and Python as the primary tools.
Twelve participants (five male) aged 25–28 years (M = 26, SD = 0.9) participated, most with prior driving and VRE experience. They preferred relaxation activities like listening to music, watching videos, or spending time in nature.

\subsubsection{Experimental Procedure}
Participants were invited to the laboratory and given a detailed introduction to the study. Following the collection of demographic information, participants watched a traffic jam video~\cite{braun2018comparison} to induce stress, establishing a baseline for the restorative VRE experience. Attention was assessed using two measures: the Attentional Function Index (AFI)\cite{cimprich2011attentional} for subjective attention evaluation and the Sustained Attention to Response Task (SART)\cite{manly2005sustained} for objective attention measurement.
The SART, implemented via an iOS app, required participants to tap for numbers (0-9) but withhold responses for the letter X and a designated control number. The system recorded accuracy (correct responses and withheld taps) and reaction time (response speed to non-targets), with higher accuracy and faster reaction times indicating better sustained attention.
After a brief training session on operating the VR system, participants engaged in a 10-minute immersive in-car VR natural environment designed to provide restorative benefits. Post-experience, participants repeated the same attention tests, with their results compared to baseline measures to evaluate the effects of the restorative experience. Finally, participants were interviewed about their feelings during the restorative experience, with their responses recorded for subsequent speech analysis. The user interviews were conducted in a semi-structured format, focusing on participants' attitudes toward the design and their overall experience. The questions covered topics such as user feedback, suggestions for improvement, and general comments on the system.
 
\subsection{User Satisfaction Baseline}

As previously mentioned, user interview content is a valuable resource for evaluating UX, particularly user satisfaction. In this study, we utilized user interview responses as the baseline for assessing satisfaction, employing advanced tools such as ELAN, OpenAI Whisper\footnote{\href{https://openai.com/index/whisper/}{https://openai.com/index/whisper/}}, and an LLM~\cite{kim2024using} to process and analyze the data comprehensively. To ensure the validity of our satisfaction measures, we cross-referenced participants' interview responses with performance metrics, including AFI scores, SART accuracy, and SART reaction times, as well as behavioral data collected during the experiment. This multi-modal approach allowed us to establish a robust baseline for user satisfaction, reducing the risk of subjective bias and enhancing the reliability of our findings.
To begin, we used ELAN and Python to isolate participants' voices from the recorded interview sessions. Since these recordings contained both the participants’ responses and the researcher’s questions, we segmented the audio to include only the participants’ speech for further analysis. The processed audio files were then transcribed into text using the OpenAI Whisper Python package, ensuring accurate capture of participants' responses for subsequent processing. The transcribed responses were analyzed using ChatGPT to perform advanced NLP and sentiment analysis~\cite{kim2024using,xing2024factors}. Each of the 12 participants’ interview transcripts was labeled (e.g., P1 to P12) for identification, and their corresponding AFI scores, SART accuracy, and SART reaction times were integrated into the analysis to provide a more comprehensive evaluation.
To conduct the analysis, the following specific prompt was used:

\begin{itemize}
\item "Based on sentiment analysis and NLP techniques, combined with participants' AFI scores, SART accuracy, and SART reaction times, classify the user satisfaction levels from the given participant responses. Provide justifications for the classification based on sentiment, emotional tone, and content relevance."
\end{itemize}

The LLM systematically analyzed the transcripts for sentiment polarity, linguistic tone, emotional expressions, and alignment with UX satisfaction themes. This process identified key themes such as enjoyment, appreciation, engagement, and constructive feedback. By correlating participants’ responses with their AFI scores, SART accuracy, and SART reaction times, the analysis provided a robust classification of user satisfaction levels. The results revealed two distinct user satisfaction groups: positive satisfaction and neutral satisfaction, with no participants classified as having negative satisfaction.
The user satisfaction groups were defined as follows:
\begin{itemize}
\item Positive Satisfaction: Participants in this group expressed clear enjoyment, strong approval, and enthusiasm about the VRE experience. Their responses were characterized by positive emotional tones, descriptions of relaxation, and explicit appreciation for the system design and functionality. These participants also showed significant improvements in AFI scores, higher SART accuracy, and faster SART reaction times, further validating their positive satisfaction.
\item Neutral Satisfaction: Participants in this group offered mixed or moderate feedback. While expressing some degree of approval, their responses often highlighted areas for improvement, such as technical challenges or mismatched expectations, and lacked strong emotional engagement with the experience. Their AFI scores showed minimal or no improvement, and their SART accuracy and reaction times remained relatively unchanged, reflecting a neutral or moderately positive attitude.
\end{itemize}

Participants in the positive satisfaction group often praised the relaxing nature of the in-car VRE environment, describing how it improved their mood and cognitive state. In contrast, participants in the neutral satisfaction group acknowledged the system's potential but pointed out limitations or suggested refinements, maintaining an overall neutral or moderately positive attitude.
This integrated methodology, combining ELAN for precise audio segmentation, OpenAI Whisper for accurate transcription, and ChatGPT for sentiment analysis, provided a scalable and efficient framework for evaluating user satisfaction. By incorporating performance metrics such as AFI scores, SART accuracy, and SART reaction times into the classification process, this approach ensures an accurate and reliable baseline for assessing user satisfaction.

\subsection{Speech Signal Analysis}
The speech data collected from participants, after being processed using ELAN and Python, was analyzed for speech signal characteristics. In this process, we explored the relationship between speech features and user satisfaction. We utilized two primary Python packages, Librosa\footnote{\href{https://librosa.org/doc/latest/index.html}{https://librosa.org/doc/latest/index.html}} and OpenSMILE\footnote{\href{https://audeering.github.io/opensmile-python/}{https://audeering.github.io/opensmile-python/}}, to extract speech features.
Specifically, we extracted both time-domain and frequency-domain speech features, including short-time energy, zero-crossing rate, jitter, and shimmer, etc. Furthermore, inspired by research on social speech dynamics~\cite{soman2010social, pentland2007social}, we also extracted social speech features such as speech activity and speech engagement. These features were analyzed to determine their relationship with user satisfaction, providing insights into how speech patterns reflect user experiences.

%% file: Sections/Results.tex
\section{Results}
To validate our approach, we extracted speech features using the Librosa and OpenSMILE Python packages. An unpaired t-test~\cite{kim2015t,boneau1960effects} revealed statistically significant differences (p < 0.05) between positive and neutral user satisfaction groups for RMS (Root Mean Square Energy), ZCR (Zero-Crossing Rate), jitter, and shimmer, highlighting their potential to differentiate between these groups. Additionally, social speech features, including activity and engagement~\cite{soman2010social, pentland2007social}, also demonstrated significant differences, with p-values below 0.05. These results emphasize the value of integrating acoustic and social speech features into user satisfaction assessment frameworks, offering a robust foundation for advancing UX evaluation methodologies.

\subsection{User Satisfaction Classification}
As presented in Table~\ref{tab:analysis} in the Appendix, the analysis of user satisfaction revealed two distinct groups: Positive Satisfaction and Neutral Satisfaction. These classifications were derived using a combination of sentiment analysis, NLP, and performance metrics, including test accuracy, reaction time (as measured by the SART test), and improvements in AFI scores. Each group’s feedback and performance metrics offer  insights into the strengths and limitations of the in-car VRE system.

\subsubsection{Positive Satisfaction}
Participants in the Positive Satisfaction group exhibited clear enjoyment and appreciation for the VR system. Their feedback often emphasized the relaxing and restorative effects of the experience. For example, P1 praised the system for its calming environment, despite raising minor concerns about falling asleep. Similarly, P8 described the experience as “very positive,” suggesting additional music options to enhance customization and enjoyment.
The performance metrics further validated this group’s satisfaction. Most participants either maintained or improved their test accuracy post-experience, reflecting enhanced focus and reduced stress. Reaction times also showed slight improvements, indicative of heightened attention. Participants like P4, with a significant score improvement of +71, highlighted the system’s ability to restore cognitive functions effectively. The Positive Satisfaction group consistently expressed enthusiasm while offering practical suggestions for enhancing the system.

\subsubsection{Neutral Satisfaction}
Participants in the Neutral Satisfaction group provided more moderate or mixed feedback. While they acknowledged some benefits of the VR system, they also highlighted areas that required improvement. For instance, P3 appreciated the system’s concept but expressed a desire for more engaging elements. Similarly, P9 found the system useful but noted discomfort with the headset and dissatisfaction with the image quality.
The performance metrics for this group presented a more varied picture. Several participants experienced declines in test accuracy post-experience (e.g., P6 and P11), and improvements in reaction time were minimal. Score improvements, while present, were generally less significant compared to the Positive Satisfaction group. For example, P9 achieved a score increase of only +24, reflecting a more limited impact of the system on their cognitive and emotional state. Participants in this group emphasized the need for technical and hardware improvements, including a more comfortable headset and enhanced system immersiveness.

\subsection{Insights from Speech Analysis Results}
\begin{figure*}[htbp]
    \centering
    \begin{subfigure}[b]{0.25\textwidth} 
        \centering
        \includegraphics[width=\textwidth]{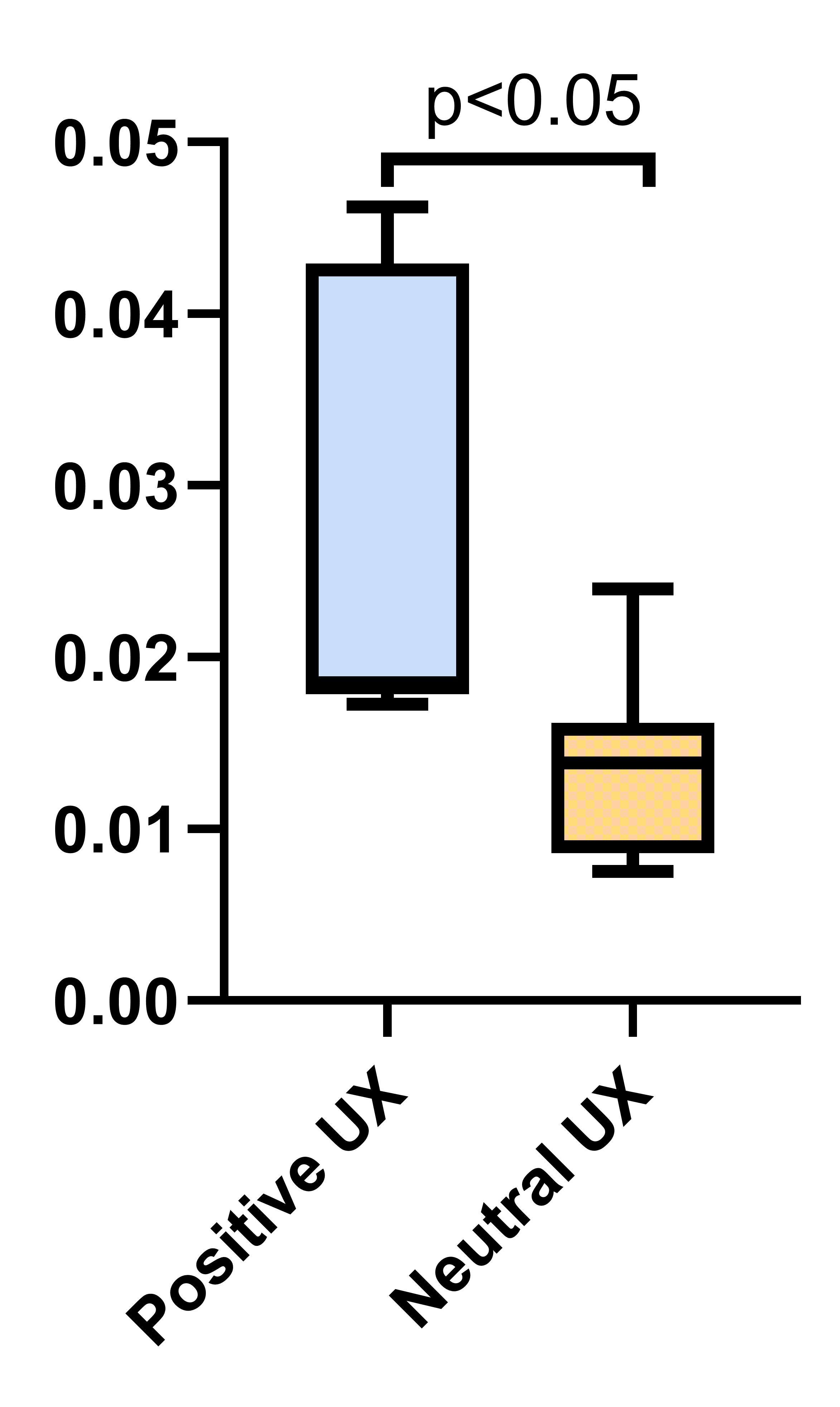}
        \caption{Root Mean Square (RMS)}
        \label{fig:RMS}
    \end{subfigure}
    \hspace{0.02\textwidth} 
    \begin{subfigure}[b]{0.25\textwidth}
        \centering
        \includegraphics[width=\textwidth]{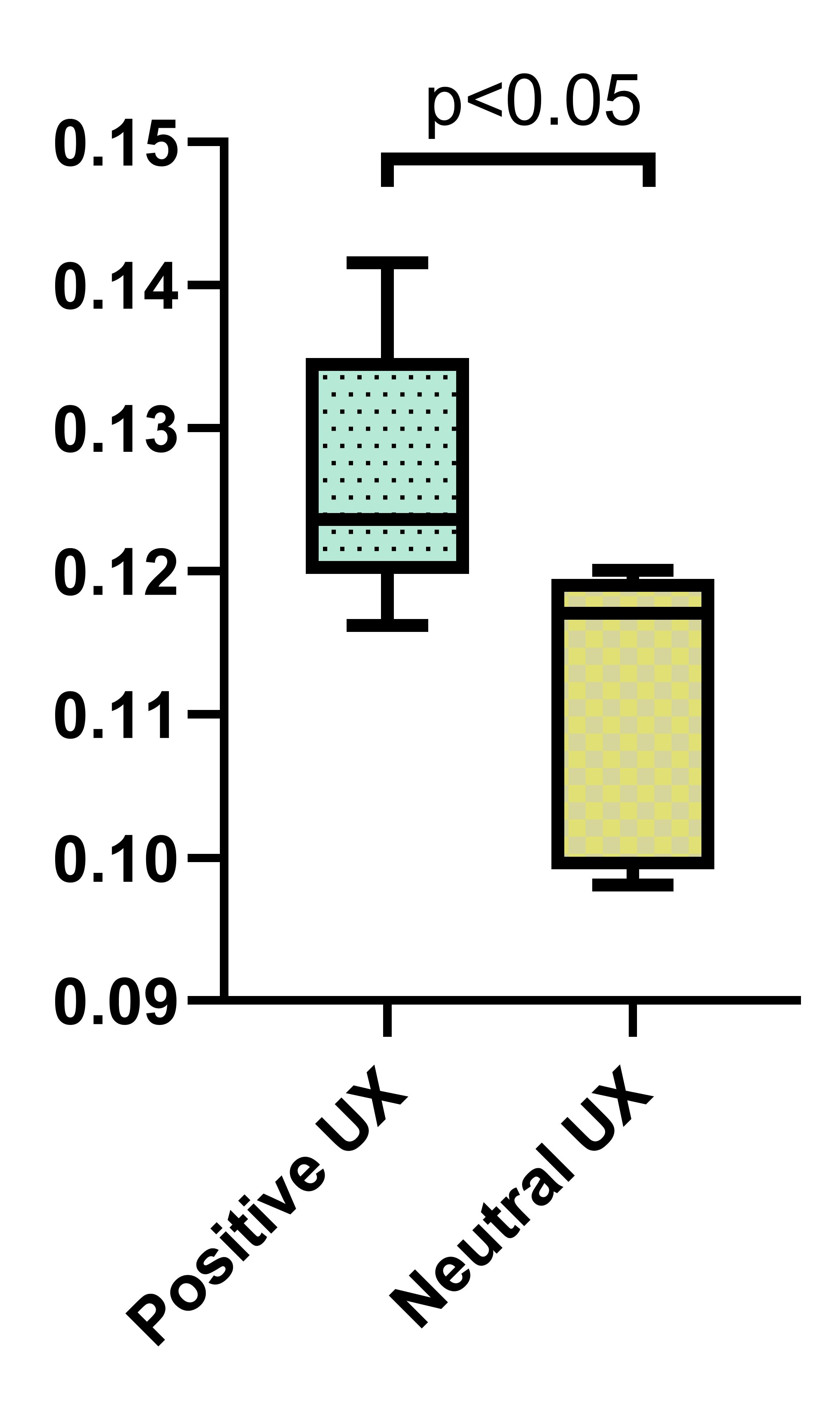}
        \caption{Zero-Crossing Rate (ZCR)}
        \label{fig:ZCR}
    \end{subfigure}
    \hspace{0.02\textwidth} 
    \begin{subfigure}[b]{0.25\textwidth}
        \centering
        \includegraphics[width=\textwidth]{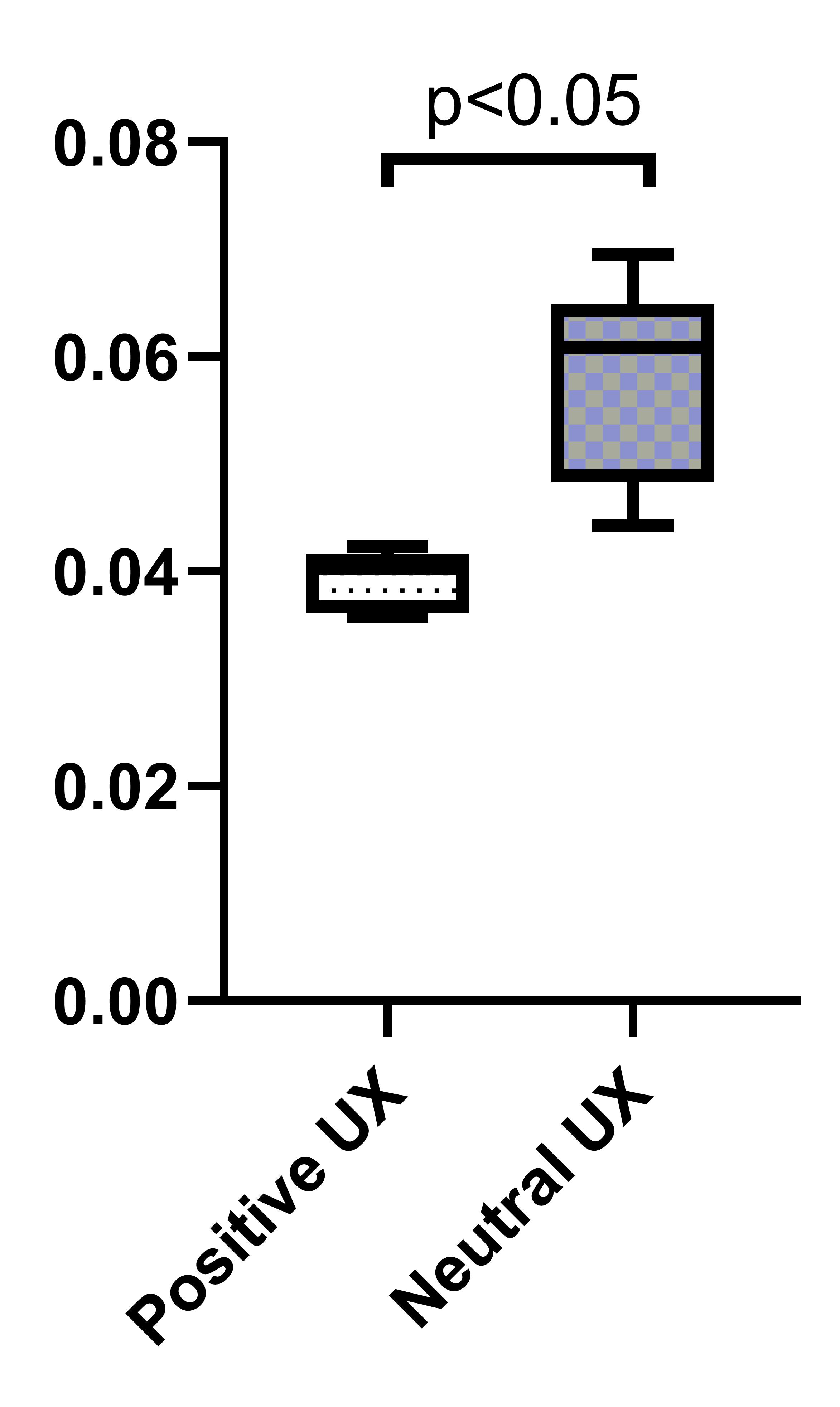}
        \caption{Speech Jitter}
        \label{fig:jitter}
    \end{subfigure}
    
    \vspace{0.3em} 
    \begin{subfigure}[b]{0.25\textwidth}
        \centering
        \includegraphics[width=\textwidth]{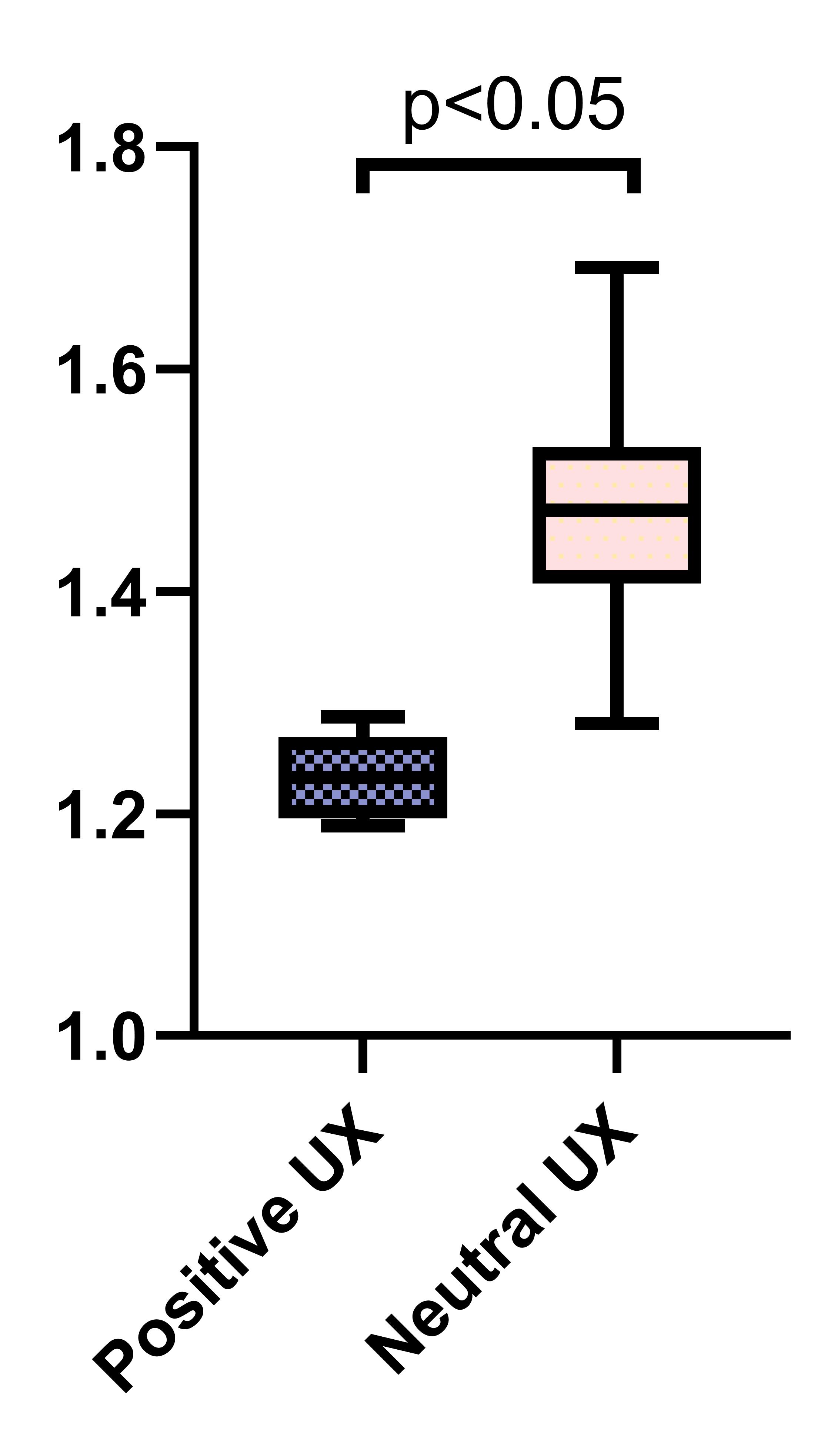}
        \caption{Speech Shimmer}
        \label{fig:shimmer}
    \end{subfigure}
    \hspace{0.02\textwidth} 
    \begin{subfigure}[b]{0.25\textwidth}
        \centering
        \includegraphics[width=\textwidth]{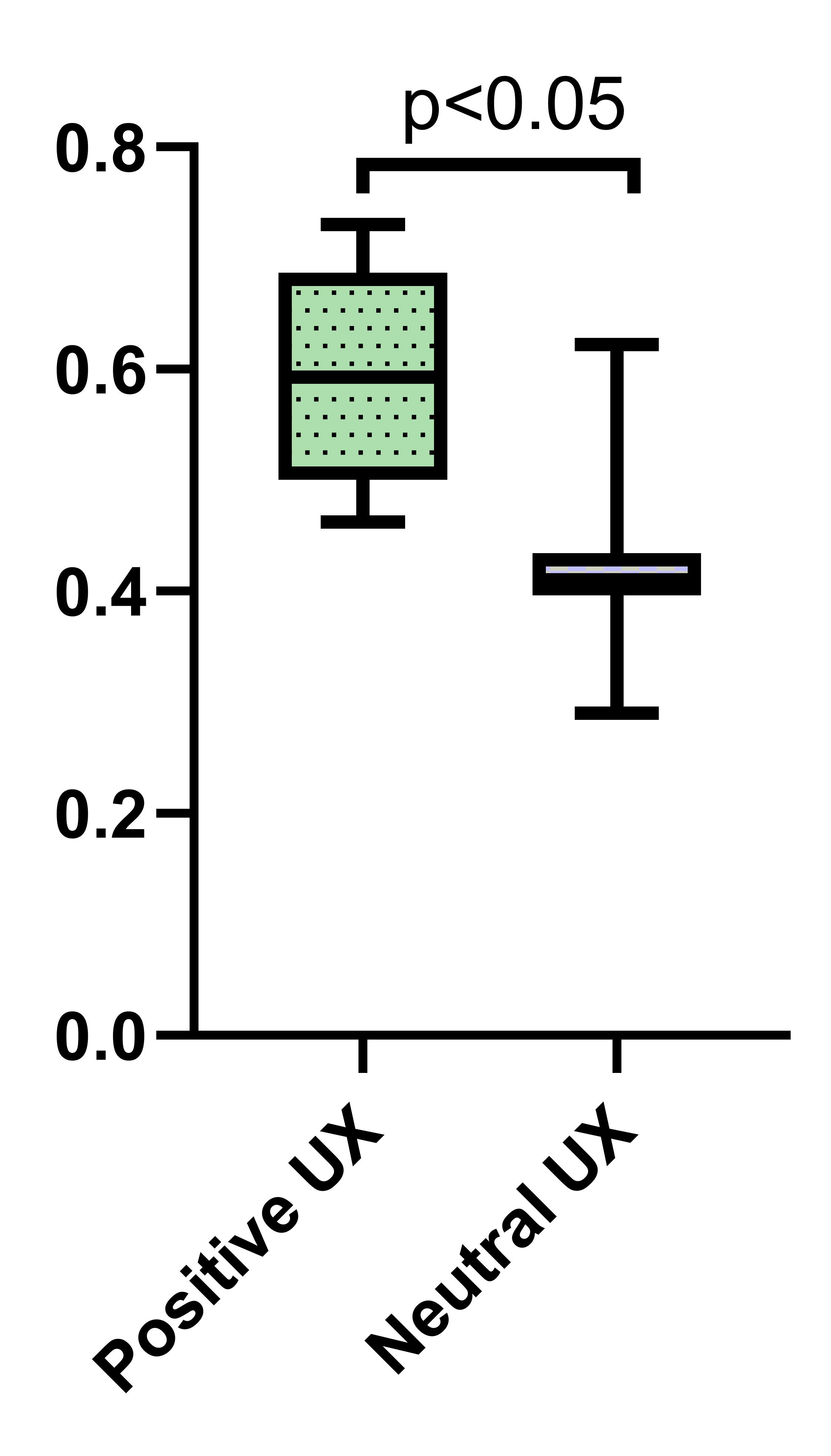}
        \caption{Speech Activity}
        \label{fig:activity}
    \end{subfigure}
    \hspace{0.02\textwidth} 
    \begin{subfigure}[b]{0.25\textwidth}
        \centering
        \includegraphics[width=\textwidth]{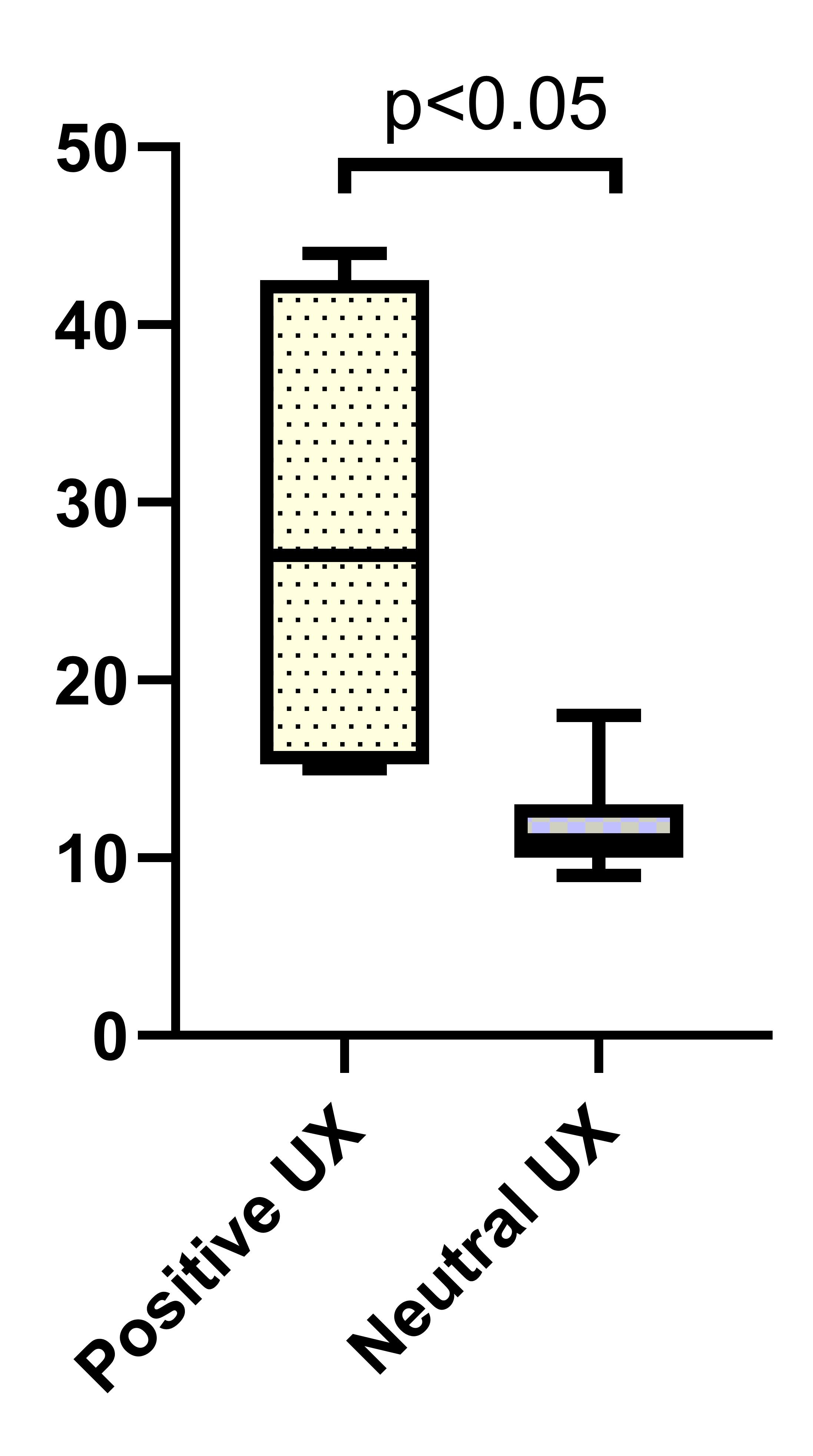}
        \caption{Speech Engagement}
        \label{fig:engagement}
    \end{subfigure}
    
    \caption{Comparison of Speech Features Between Two User Satisfaction Groups Using T-Test. The figures include both acoustic features (RMS, ZCR, Jitter, Shimmer) and social speech features (Activity, Engagement).}
    \label{fig:3x2grid_small}
\end{figure*}

\paragraph{\textbf{Time and Frequency Domain Speech Feature}}
To extract time and frequency domain speech features, we utilized Librosa and OpenSMILE to extract as many relevant features as possible, as described earlier. As shown in Figure~\ref{fig:RMS} and Figure~\ref{fig:ZCR}, we observed that RMS and ZCR demonstrated significant differences between the two user satisfaction groups, with p-values of 0.0333 and 0.0199, respectively.
RMS measures the average loudness of an audio waveform~\cite{glasberg2002model}, while ZCR is commonly used to distinguish between voiced and unvoiced sounds~\cite{bachu2010voiced}. These findings suggest that both RMS and ZCR can effectively differentiate between the two user satisfaction groups.
Additionally, as illustrated in Figure~\ref{fig:jitter} and Figure~\ref{fig:shimmer}, jitter and shimmer also exhibited significant differences between the user satisfaction groups, with p-values of 0.0017 and 0.0018, respectively. In speech processing, jitter refers to the variability or perturbation of the fundamental frequency, while shimmer reflects similar perturbations in the amplitude of the sound wave~\cite{wertzner2005analysis}. These results further indicate that both jitter and shimmer are valuable features for distinguishing between different levels of user satisfaction.

\paragraph{\textbf{Social Speech Features}}

The Figure~\ref{fig:activity} and Figure~\ref{fig:engagement} illustrate the differences in speech activity and speech engagement between the positive and neutral user satisfaction groups, as determined through statistical analysis.
In Figure~\ref{fig:activity}, the speech activity levels are significantly higher in the Positive UX group compared to the Neutral UX group, with a p-value of 0.0172. This indicates that participants with positive satisfaction were more actively involved during the interaction, speaking more frequently or for longer durations in the interviews. This finding suggests that higher speech activity often reflects greater enthusiasm or engagement in the interaction, traits typically associated with positive user satisfaction.
Similarly, in Figure~\ref{fig:engagement}, speech engagement, which measures how a speaker's contributions influence the flow of conversation, also demonstrates a statistically significant difference between the two groups, with a p-value of 0.0177. The Positive UX group shows higher engagement levels, indicating that participants in this group contributed more dynamically to the conversations, fostering smoother and more interactive exchanges. Conversely, the Neutral UX group displayed lower engagement, suggesting a more passive interaction style.
These findings highlight the importance of speech activity and speech engagement as valuable social speech features for differentiating user satisfaction levels and enriching the assessment of user experience.

\paragraph{\textbf{User Satisfaction Level Classification}}
To evaluate the effectiveness of the six speech features correlated with user satisfaction levels, we applied traditional machine learning methods, including Support Vector Machine (SVM) and k-Nearest Neighbors (KNN), for classification. The results demonstrated that the SVM algorithm achieved an accuracy of 86\%, while the KNN algorithm achieved an accuracy of 75\% in distinguishing between the two user satisfaction groups.
These results indicate that the selected speech features are not only effective in differentiating between positive and neutral user satisfaction groups but also hold promise for broader applications in user satisfaction detection. This underscores the potential of integrating these features into objective UX evaluation frameworks. Furthermore, these features could be leveraged in future studies or practical implementations to enhance the accuracy and reliability of user satisfaction analysis, advancing the methodologies for evaluating user experience.

%% file: Sections/Discussion.tex
\section{Discussion}
The findings of this study underscore the potential of utilizing speech features as objective indicators for evaluating user satisfaction. By integrating time-domain, frequency-domain, and social speech features, we demonstrated their statistically significant ability to distinguish between positive and neutral user satisfaction groups. These results contribute to the advancement of HCI research by introducing a novel approach to UX evaluation. This approach reduces reliance on subjective methods, such as questionnaires and self-reported metrics, which are often prone to subjective bias, thereby enhancing the objectivity and reliability of UX assessments.

\subsection{Speech-Based UX Evaluation and Key Findings}
The findings of this study highlight the potential of speech features as objective indicators for evaluating user satisfaction. By integrating time-domain, frequency-domain, and social speech features, we demonstrated their ability to distinguish between satisfaction groups with statistical significance. This approach reduces reliance on subjective self-reported methods, enhancing the objectivity and reliability of UX assessments in HCI research. Our analysis confirmed that features such as RMS, ZCR, jitter, and shimmer played a significant role in differentiating user satisfaction levels. Additionally, social speech features—such as speech activity and engagement—were effective in capturing variations in user interaction quality. These results indicate that speech-driven UX evaluation can provide real-time insights for interactive systems, making it a valuable tool for adaptive interface design and user-centered improvements.

\subsection{Generalization Beyond the In-Car VRE Context}
While this study focused on an in-car VRE environment, its findings have broader implications for UX evaluation in other interactive systems. Speech-based UX assessment can be applied to voice-assisted AI interactions, virtual reality experiences, and digital collaboration tools. For example, in voice assistants, speech engagement features can enhance system responsiveness, while in VR environments~\cite{zadeh2023adaptive}, speech prosody can provide insights into user immersion and cognitive load~\cite{li2021journey,ma2021you}. Similarly, in collaborative tools, social speech features~\cite{pentland2010honest} can be used to analyze group dynamics and teamwork effectiveness. Expanding the scope of speech-based UX evaluation to these domains would enable a more comprehensive understanding of user satisfaction in different contexts. Future studies should explore the applicability of these features in real-time interactive settings, integrating multimodal data sources—such as facial expressions and physiological responses—to enrich UX evaluation frameworks.

\subsection{Addressing Confounding Factors in UX Evaluation}
A key consideration in this study was the potential confounding of linguistic elements (e.g., speech features) and satisfaction measures. To mitigate this, we carefully controlled for external factors that could influence speech patterns, such as background noise, interviewer tone, and participants' speaking habits. Additionally, we analyzed the correlation between speech features and performance metrics (AFI and SART) to ensure that the observed differences in speech patterns were indeed reflective of user satisfaction and not influenced by other variables. This approach strengthens the validity of our findings and ensures that the relationship between speech features and user satisfaction is not confounded by external factors.

\subsection{Limitations and Future Work}
Despite the promising findings, this study has several limitations that should be addressed in future research. First, the small sample size (N=12) limits generalizability, as larger and more diverse participant groups are needed to validate these results across different demographics and interaction settings. Future research should incorporate participants from varied age groups, cultural backgrounds, and UX familiarity levels to ensure robustness in speech-based UX evaluation. Second, the binary classification approach (positive vs. neutral) simplifies the complexity of user experiences. While this method offers interpretability, user satisfaction is inherently multidimensional. Future research should explore multi-class classification models, regression-based UX scoring, or continuous satisfaction scales to capture a more granular understanding of UX. Third, while we analyzed key speech features (RMS, ZCR, jitter, shimmer, engagement, and activity), additional prosodic and spectral features can be further examined. These features may play a more significant role in different UX contexts, particularly those involving emotional arousal and cognitive demand. Future research should leverage larger datasets and deep learning-based models to refine feature selection and enhance predictive accuracy. Lastly, integrating real-time speech monitoring and multimodal data sources (e.g., physiological signals, gaze tracking) will further improve UX assessment accuracy. Implementing dynamic speech-based UX feedback systems could enable adaptive interface adjustments based on real-time user responses, improving overall usability and engagement. By addressing these challenges, future research can further strengthen the scalability and applicability of speech-driven UX assessment across diverse HCI domains.


%% file: Sections/Conclusion.tex
\section{Conclusion}
This study demonstrates the feasibility of using speech features as objective indicators for UX evaluation. By analyzing features such as RMS, ZCR, jitter, shimmer, speech activity, and engagement, we successfully differentiated positive and neutral satisfaction groups with statistical significance. Machine learning models like SVM and KNN further validated the approach with high classification accuracy. This work offers a scalable, reliable framework for UX evaluation, paving the way for integrating speech analysis into HCI research and design.
Future research should expand on these findings by employing larger, more diverse datasets, exploring advanced machine learning models, and incorporating multimodal data, such as facial expressions and physiological signals, to enhance the generalizability, robustness, and practical applicability of this approach in more UX evaluation scenarios.